\documentclass[sigconf]{acmart}
\usepackage{amsmath}
\usepackage{color}
\usepackage{subfigure}
\usepackage{array}
\usepackage{{inputenc}}
\usepackage{balance}
\usepackage{multirow,tabularx}
\usepackage{booktabs,longtable}
\usepackage{algorithm}
\usepackage{algorithmic}
\usepackage{graphicx}
\usepackage{enumitem}
\usepackage{verbatim}
\usepackage{amsfonts}
\usepackage{hyperref}
\hypersetup{colorlinks=false,linkcolor=blue,urlcolor=blue,citecolor=red}
\newcommand{\Space}[1]{\mathbb{#1}}
\newcommand{\Set}[1]{\mathcal{#1}}

\newcommand{\ie}{\emph{i.e., }}
\newcommand{\eg}{\emph{e.g., }}

\newcommand{\wrt}{\emph{w.r.t. }}

\newcommand{\aka}{\emph{aka. }}
\newcommand{\argmin}{\mathop{\mathrm{argmin}}}

\AtBeginDocument{%
  \providecommand\BibTeX{{%
    \normalfont B\kern-0.5em{\scshape i\kern-0.25em b}\kern-0.8em\TeX}}}

\copyrightyear{2022}
\acmYear{2022}
\setcopyright{acmcopyright}\acmConference[SIGIR '22]{Proceedings of the 45th International ACM SIGIR Conference on Research and Development in Information Retrieval}{July 11--15, 2022}{Madrid, Spain}
\acmBooktitle{Proceedings of the 45th International ACM SIGIR Conference on Research and Development in Information Retrieval (SIGIR '22), July 11--15, 2022, Madrid, Spain}
\acmPrice{15.00}
\acmDOI{10.1145/3477495.3532059}
\acmISBN{978-1-4503-8732-3/22/07}
\begin{document}
% remove the headers
\fancyhead{}

% multi-facet
\title{Self-Guided Learning to Denoise for Robust Recommendation}

%  \author{Yunjun Gao$^1$, Yuntao Du$^1$, Yujia Hu$^1$, Lu Chen$^{1*}$, Xinjun Zhu$^2$, Ziquan Fang$^1$ and Baihua Zheng$^3$}
 
%  \def \authors{Yunjun Gao, Yuntao Du, Yujia Hu, Lu Chen, Xinjun Zhu, Ziquan Fang and Baihua Zheng}
 
% \affiliation{
%  {\large$^1$}College of Computer Science, Zhejiang University, Hangzhou, China\\
%  {\large$^2$}School of Software, Zhejiang University, Ningbo, China\\
%  {\large$^3$}School of Computing and Information Systems, Singapore Management University, Singapore
%  \country{}}
 
%  \email{{gaoyj,ytdu,yjhu,luchen,xjzhu,zqfang}@zju.edu.cn, bhzheng@smu.edu.sg}

% \thanks{$^*$Lu Chen is the corresponding author.}

\author{Yunjun Gao, Yuntao Du, Yujia Hu, \\Lu Chen, Xinjun Zhu, Ziquan Fang}
\affiliation{%
    \institution{College of Computer Science, Zhejiang University, China}
    \country{}
}
\email{{gaoyj, ytdu, yjhu, luchen, xjzhu, zqfang}@zju.edu.cn}

% \author{Yunjun Gao}
% \affiliation{%
%     \institution{College of Computer Science, Zhejiang University}
%     \country{}
% }
% \email{gaoyj@zju.edu.cn}

% \author{Yuntao Du}
% \affiliation{%
%     \institution{College of Computer Science, Zhejiang University}
%     \country{}
% }
% \email{ytdu@zju.edu.cn}

% \author{Yujia Hu}
% \affiliation{%
%     \institution{College of Computer Science, Zhejiang University}
%     \country{}
% }
% \email{yjhu@zju.edu.cn}

% \author{Lu Chen}
% \authornote{Lu Chen is the corresponding author.}
% \affiliation{%
%     \institution{College of Computer Science, Zhejiang University}
%     \country{}
% }
% \email{luchen@zju.edu.cn}

% \author{Xinjun Zhu}
% \affiliation{%
%     \institution{School of Software Technology, Zhejiang University}
%     \country{}
% }
% \email{xjzhu@zju.edu.cn}

% \author{Ziquan Fang}
% \affiliation{%
%     \institution{College of Computer Science, Zhejiang University}
%     \country{}
% }
% \email{zqfang@zju.edu.cn}

\author{Baihua Zheng}
\affiliation{%
    \institution{School of Computing and Information Systems,\\ Singapore Management University, Singapore}
    \country{}
}
\email{bhzheng@smu.edu.sg}

\renewcommand{\shortauthors}{Gao, et al.}

%%
%% The abstract is a short summary of the work to be presented in the
%% article.
\begin{abstract}
The ubiquity of implicit feedback makes them the default choice to build modern recommender systems. Generally speaking, observed interactions are considered as positive samples, while unobserved interactions are considered as negative ones.
However, implicit feedback is inherently noisy because of the ubiquitous presence of \textit{noisy-positive} and \textit{noisy-negative} interactions. Recently, some studies have noticed the importance of denoising implicit feedback for recommendations, and enhanced the robustness of recommendation models to some extent. Nonetheless, they typically fail to (1) capture the hard yet clean interactions for learning comprehensive user preference, and (2) provide a universal denoising solution that can be applied to various kinds of recommendation models.

In this paper, we thoroughly investigate the memorization effect of recommendation models, and propose a new denoising paradigm, \ie \underline{S}elf-\underline{G}uided \underline{D}enoising \underline{L}earning (SGDL), which is able to collect memorized interactions at the early stage of the training (\ie ``noise-resistant'' period), and leverage those data as denoising signals to guide the following training (\ie ``noise-sensitive'' period) of the model in a meta-learning manner. Besides, our method can automatically switch its learning phase at the memorization point from memorization to self-guided learning, and select clean and informative memorized data via a novel adaptive denoising scheduler to improve the robustness. We incorporate SGDL with four representative recommendation models (\ie NeuMF, CDAE, NGCF and LightGCN) and different loss functions (\ie binary cross-entropy and BPR loss). The experimental results on three benchmark datasets demonstrate the effectiveness of SGDL over the state-of-the-art denoising methods like T-CE, IR, DeCA, and even state-of-the-art robust graph-based methods like SGCN and SGL.

\end{abstract}

%%
%% The code below is generated by the tool at http://dl.acm.org/ccs.cfm.
%% Please copy and paste the code instead of the example below.
%%


%% Keywords. The author(s) should pick words that accurately describe
%% the work being presented. Separate the keywords with commas.

\maketitle

\section{Introduction}
\label{sec:introduction}

\begin{figure*}[t]
    \vspace{-5pt}
	\centering
	\includegraphics[width=0.39\textwidth]{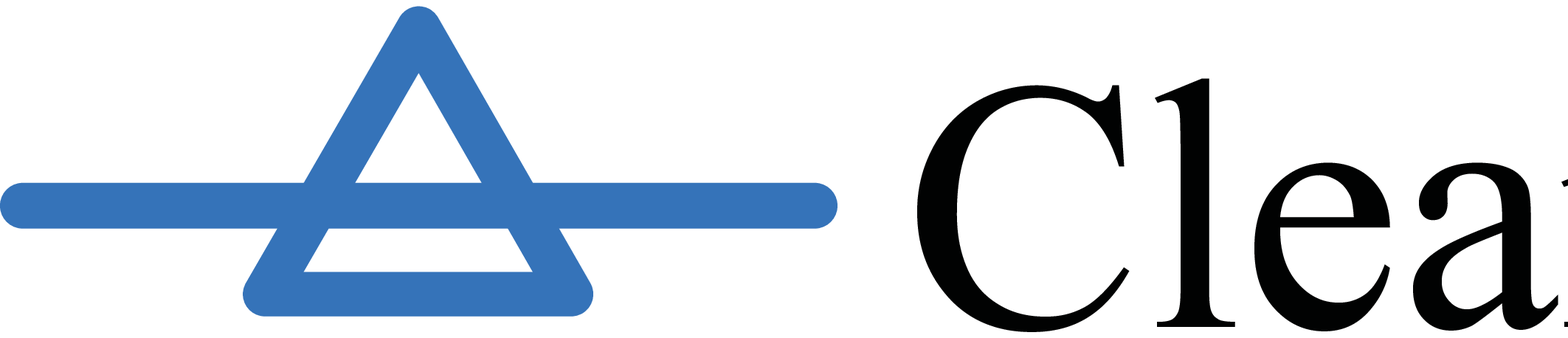}
	\vspace{3pt}
	\includegraphics[width=1.0\textwidth]{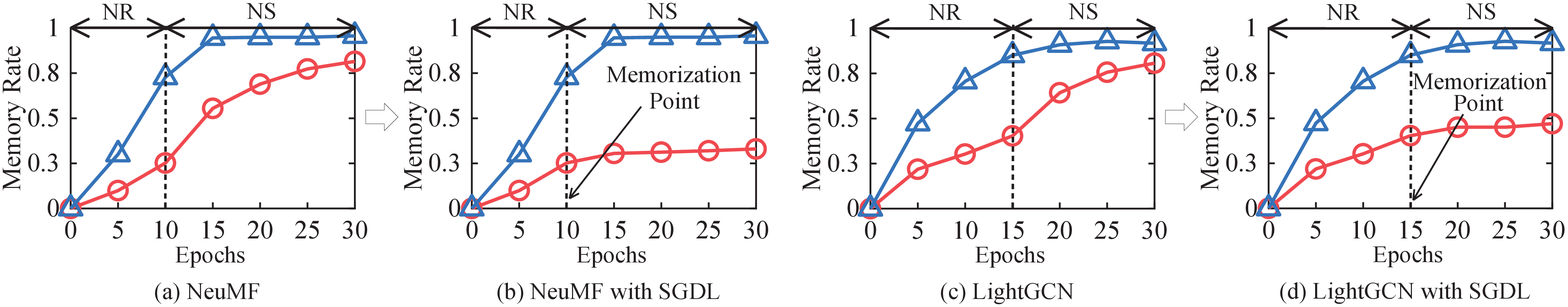}
	\vspace{-20pt}
	\caption{Key idea of SGDL: (a) and (c) show the memory rate when training NeuMF and LightGCN on the MovieLens dataset, respetively; (b) and (d) show the memory rate when training the corresponding model with SGDL. The memory rate is the proportion of memorized data (see Section~\ref{sec:memorized-data} for details) in the clean and noisy interactions. NR denotes ``noise-resistant'' period and NS represents ``noise-sensitive'' period, respectively.}
	\label{fig:memorization}
	\vspace{-10pt}
\end{figure*}

Recommender systems have been widely deployed to perform personalized information
filtering, especially for various online services such as E-commerce~\cite{sigir19e-commerce,gharibshah2020deep}, social media~\cite{wsdm17socialRS}, and news portals~\cite{www18dkn}. Most of existing recommender systems use implicit feedback (\eg view and click) to develop machine learning models, due to its large volume and availability~\cite{uai09BPRloss,sigir16implicit-fm,tkde22metakg,sigir22HAKG}. Specifically, the observed interactions between users and items are viewed as positive instances, while unobserved interactions are viewed as negative instances. However, implicit feedback is inherently noisy because of the ubiquitous presence of \textit{noisy-positive} and \textit{noisy-negative} interactions~\cite{cidm08noise,recsys21denoise,mm21occf,sigir21boostrapping}. Take the E-commerce scenario as an example. A large portion of click behaviors are triggered by the curiosity of users, which cannot directly indicate the users' positive views of the products. On the other hand, unobserved interactions may attribute to the unawareness of users because the items are simply not exposed to them. Hence, blindly fitting the implicit feedback to recommender systems without considering the inherent noise would fail to understand users' true preferences, and eventually harm user experiences and degrade recommendation performance~\cite{ijcai20feedback,wsdm21denoise}.

Considering the widespread use of implicit feedback and its large impact on the recommendation model, some recent studies have noticed the importance of denoising implicit feedback for recommendations. Existing efforts on tackling this problem can be roughly divided into two categories: \emph{sample selection methods}~\cite{kdd12ranking,tkde19sampling-bpr,sigir20sampler,mm21occf} and \emph{sample re-weighting methods}~\cite{wsdm21denoise,cikm21sequential-denoise,arxiv22ensemble}. Sample selection methods focus on designing more effective samplers to collect clean samples for learning users' preferences, while their performance suffers from high variance since they heavily depend on the sampling distribution~\cite{uai18fbgd}. On the other hand, sample re-weighting methods aim to distinguish noisy interactions from clean data in terms of the loss values, and assign lower weights to noisy interactions with high loss values during training. Their key idea is consistent with \textit{memorization effect}~\cite{icml17memory}: models tend to initially learn easy and clean patterns (\ie user preferences) in the early stage of their learning, and eventually memorize all training interactions. Benefiting from this principle, sample re-weighting methods are able to successfully identify noisy interactions.

Although sample re-weighting methods can achieve promising performance for denoising implicit feedback, we want to highlight that they commonly suffer from following \textit{two} problems:

% limitations
\begin{itemize}[leftmargin=*]
    \item \textbf{Abandon of Hard Clean Interactions.} These methods heavily rely on the loss values. They simply assume that interactions with large loss values are noisy and penalize them with small weights. However, it has been reported that some clean interactions (\aka hard yet clean interactions) may also have high loss values at the beginning of the training, and those interactions play an important role in understanding users' behaviors and preferences~\cite{ijcai19sampling,cikm21sequential-denoise,arxiv22ensemble}. Nonetheless, these interactions are simply discarded by existing methods due to high loss values, incurring insufficient understanding of users' true preferences.
    
    \item \textbf{Lack of Adaptivity and Universality.} Existing sample re-weighting methods are able to achieve good performance \emph{only} when the reweighting configurations (e.g., weights and thresholds) are properly specified. However, obtaining proper configurations normally requires time-consuming procedures (\eg grid search), and the optimal configurations for one dataset typically are not applicable for other datasets because of the different data distributions. Meanwhile, these methods are only applicable to the recommender models with predefined pointwise loss function (\eg cross-entropy loss in~\cite{wsdm21denoise,arxiv22ensemble}), which makes it hard to incorporate other popular ranking loss functions (\eg BPR loss~\cite{uai09BPRloss}) and limits their applications.
\end{itemize}

% our solution
In this regard, we have thoroughly explored the memorization of the representative recommendation models (\eg NeuMF~\cite{www17nfm} and LightGCN~\cite{sigir20lighGCN}) on implicit feedback, including the clean interactions and the noisy interactions. By analyzing the learning processes of different models, we have observed the existence of two learning periods. As shown in Figure~\ref{fig:memorization}, a typical learning process consists of the ``noise-resistant'' (NR) period and the ``noise-sensitive'' (NS) period. The former is the duration when the memorization of noisy interactions is insignificant because the models focus on memorizing easy and clean patterns at their early stage of training; while the latter is the duration when the memorization of noisy interactions rapidly increases since models eventually begin memorizing all the implicit feedback at the late stage of training. The timestamp in the training process that can best differentiate above two periods is defined as \textit{memorization point} of the model training, which are represented by dotted lines in Figure~\ref{fig:memorization}. 

These observations motivate us to design a new approach that better leverages the memorization nature of recommendation models. Specifically, we propose a new denoising paradigm, namely, \underline{S}elf-\underline{G}uided \underline{D}enoising \underline{L}earning (SGDL), which is able to collect memorized interactions at the early stage of training (\ie ``noise-resistant'' period), and leverage those data as denoising signals to guide the subsequent training (\ie ``noise-sensitive'' period) of the model in a meta-learning manner. Besides, SGDL can automatically transit from noise-resistant period to noise-sensitive period in order to stop accumulating memorized interactions at the \textit{memorization point}, since the noisy interactions are gradually memorized with the training process. In a nutshell, corresponding to the two observed periods commonly existing in the training process of different models, SGDL contains two key phases, \ie \textit{memorization} and \textit{self-guided learning}:

\begin{itemize}[leftmargin=*]
    \item \textbf{Memorization.} Owing to the negligible memorization of noisy interactions during  ``noise-resistant'' period, the model is initially trained with all the implicit feedback. To better reveal the underlying memorization nature of the model during training, we design the new memorization-based metrics to define the memorization states of data. As the memorized data are mostly easy and clean interactions at the early stage of training, we collect them as denoising signals to guide the following denoising training process. Moreover, SGDL is able to automatically estimate the best memorization point, from which the learning moves from the ``noise-resistant'' period to ``noise-sensitive'' period, to stop accumulating memorized data without \textit{any} supervision.
    \item \textbf{Self-Guided Learning.} To avoid the memorization of noisy interactions in the ``noise-sensitive'' period, we leverage the memorized data collected from the ``noise-resistant'' period to represent user preferences. Specifically, a denoising module is proposed to learn a parameterized weighting function for the implicit feedback, which is guided by memorized data and updated with the learning process of the model. Moreover, since some of the memorized data can also be noisy, we further develop a novel adaptive denoising scheduler to prevent the denoising module from being corrupted by noisy yet memorized samples. Technically, the adaptive denoising scheduler characterizes the contribution of each memorized data to the denoising performance, and decides whether to use the sample by predicting its probability being sampled. The scheduler is also simultaneously optimized together with the learning process to enhance the robustness of the model.
\end{itemize}

Through the above two phases, SGDL can denoise implicit feedback with the help of memorized data, which are naturally collected by exploiting the noise-resistant period of training. Compared with standard training, SGDL can dramatically reduce the memory ratio of noisy samples in the noise-sensitive period, and enhance the robustness of recommendation models, as shown in Figure~\ref{fig:memorization}. Moreover, since the model is constantly trained with all implicit feedback data, our method can help the model learn users' true preferences with hard yet clean samples, leading to better recommendation performance. Last but not the least, our method does not need \textit{any} thresholds or predefined weighting functions, and is easy to be applied to \textit{any} learning-based recommendation models. We conduct extensive experiments on three real-world datasets with four representative recommendation models (\ie NeuMF~\cite{www17nfm}, CDAE~\cite{wsdm16cdae}, NGCF~\cite{sigir19ngcf}, and LightGCN~\cite{sigir20lighGCN}) and two loss functions (\ie binary cross-entropy loss and BPR loss~\cite{uai09BPRloss}). Experimental results show that our SGDL significantly outperforms all state-of-the-art denoising methods, and it achieves comparable (in many cases even better) performance to the state-of-the-art robust graph-based methods like SGCN~\cite{sigir21mask} and SGL~\cite{sigir21sgl}.

In summary, we make three key contributions in this paper, as listed below.

\begin{itemize}[leftmargin=*]
    \item We develop a new denoising paradigm, \ie \emph{self-guided denoising learning (SGDL)}, which leverages the self-labeled memorized data as guidance to offer denoising signals for robust recommendation without defining \textit{any}  weighting functions or requiring \textit{any} auxiliary information.
    \item We carefully exploit the memorization effect of recommendation models, and design two training phases that can collect memorized data and utilize them as guidance to denoise implicit feedback, respectively. Besides, a novel adaptive denoising scheduler is introduced to further improve the robustness.
    \item We incorporate SGDL with four representative recommendation models, and  conduct extensive experiments on three public benchmark datasets with various state-of-the-art methods to demonstrate the superiority and universality of SGDL.
\end{itemize}

\section{Problem Formulation}
\label{sec:problem_formulation}
We first introduce the common paradigm of user preferences learning from implicit feedback, and then formulate our task.

\vspace{5pt}
\noindent
\textbf{Preference learning from implicit feedback.} In this paper, we focus on learning the user preferences from the implicit feedback~\cite{uai09BPRloss}. Specifically, the behavior data (\eg click and review) $\Set{D}=\{u,i,y_{ui}|u\in\Set{U},i\in\Set{I}\}$ involves a set of users $\Set{U} = \{u\}$ and a set of items $\Set{I} = \{i\}$, as well as the interaction $y_{ui} = \{0,1\}$ that indicates whether user $u$ has interacted with item $i$. Most of the state-of-the-art recommendation methods (\eg NeuMF~\cite{www17nfm} and LightGCN~\cite{sigir20lighGCN}) assume that the interaction $y_{ui}$ could represent the users' true preferences, and directly learn the model $f$ with parameters $\theta$ by minimizing a ranking loss function over $\Set{D}$.

\vspace{5pt}
\noindent
\textbf{Denoising implicit feedback for recommendations.} However, due to the existence of inherent noise in implicit feedback, recommendation models might fail to learn the users' true preferences with typical training process, resulting in suboptimal performance. Thus, the task of the paper is, given the noisy implicit feedback $\Set{D}$ that contains both noisy-positive and noisy-negative feedback, to infer users' true preferences with optimal model parameters $\theta^*$. 

\begin{figure}[t]
    \centering
	\includegraphics[width=0.49\textwidth]{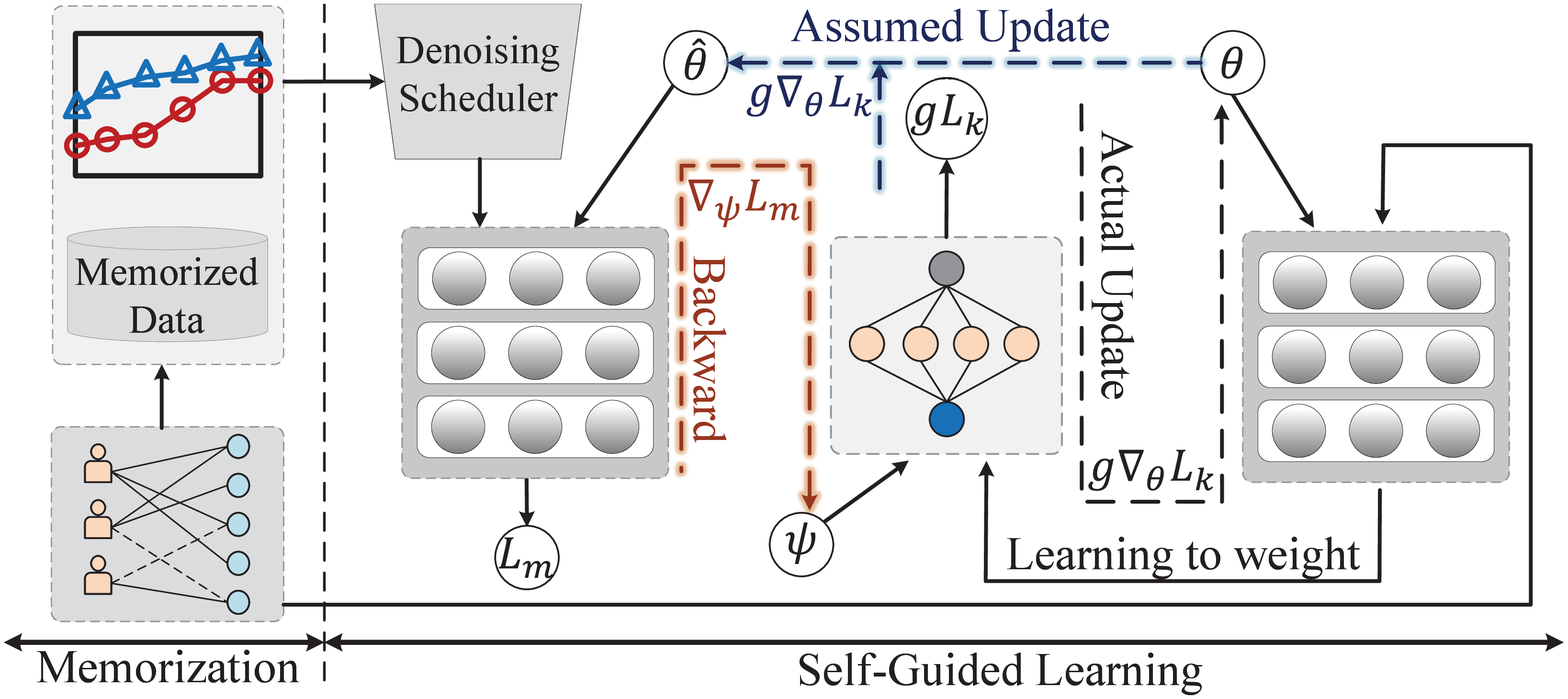}
	\vspace{-20pt}
	\caption{The overall framework of SGDL. During Phase I (memorization), memorized interactions are collected as denoising signals; During Phase II (self-guided learning), the weighting function is simultaneously learned with the recommender model, which is guided by the memorization data in a meta-learning manner (see Section~\ref{sec:denoising_learning} for details).}
	\label{fig:framework}
	\vspace{-10pt}
\end{figure}

\section{Methodology}
\label{sec:methodology}

In this section, we detail SGDL that comprises the following two phases: (i) \emph{memorization}, which exploits the memorization effect of models to collect memorized data during the noise-resistant period and estimates the best memorization point to automatically transit to phase II; and (ii) \emph{self-guided learning}, which leverages the memorized samples as denoise signals to guide the model learning during the noise-sensitive period, and discards the potential noisy interactions in memorized data with a novel adaptive denoising scheduler for robustness. We detail the two phases as follows.

\subsection{Phase I: Memorization}
\label{sec:phase1}

Initially, Phase I tries to train the recommendation model in a conventional way during the noise-resistant period, where the memorization of noisy interactions is assumed to be suppressed, as we discover in Figure~\ref{fig:memorization}. Since most of the memorized interactions until the memorization point are clean, we collect them to form the memorized data, which can be used as the denoising signals to guide the training in Phase II. Therefore, the major challenges in Phase I include 1) how to define the memorization of interactions and 2) how to estimate the memorization point.

\subsubsection{\textbf{Memorized Interactions.}}
\label{sec:memorized-data}
Previous studies~\cite{wsdm21denoise, arxiv22ensemble} mainly use loss values of training data to demonstrate the memorization effect of recommendation models. However, we argue that loss values are insufficient to reflect the learning process, since they are inconsistent with the optimization target of recommendation models (\ie personalized ranking), and unable to distinguish hard interactions from noisy ones. Thus, it is necessary to design a new memorization-based metric, which takes the learning process of recommendation models into consideration.

Inspired by the widely used hit ratio metric~\cite{uai09BPRloss,recsys10metric}, we define that an interaction $(u,i)$ is \textit{memorized} at epoch $t$ by the model $\theta$ if item $i$ is in the ranking list of user $u$, denoted as $m_t(u,i)$. To ensure the reliability of ranking results, for each user $u$, we include the top-$N$ of the ranking items into its ranking list,
%to evaluate,
where $N$ is the length of observed interactions. However, simply calculating the memorization of interactions of a single epoch could lead to unstable results, since the model is not well trained during the noise-resistant period. Hence, we trace the memorization states of interactions for the most recent $h$ epochs, and define the final memorization of interaction $(u,i)$ as follows:
\begin{gather}\label{equ:memorized_interaction}
m_t^h(u,i) = \frac{1}{{|\Set{P}_t^h(u,i)|}}\sum\nolimits_{m_j(u,i)\in\Set{P}_t^h(u,i)}m_i(u,i)
\end{gather}
where $\Set{P}_t^h(u,i)=\{m_{t-h+1}(u,i),\cdots,m_{t}(u,i)\}$ captures the most recent $h$ memorization histories of interaction $(u,i)$. We define an interaction $(u,i)$ \textit{memorized} by a model $\theta$ if the majority of the recent histories $\Set{P}_t^h(u,i)$ coincides with the memorization state (\ie $m_t^h(u,i)$ is larger than $0.5$). It is worth mentioning that the definition of memorized interactions dose not need \textit{any} labels for supervision, and is able to effectively indicate the underlying memorization effect of recommendation models.

\begin{figure}[t]
    \centering
    \includegraphics[width=0.2\textwidth]{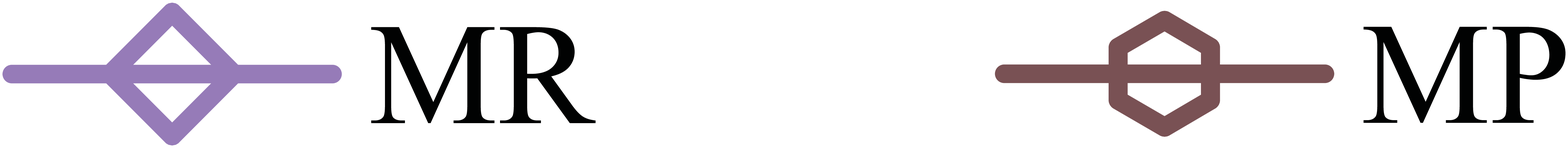}
	\includegraphics[width=0.48\textwidth]{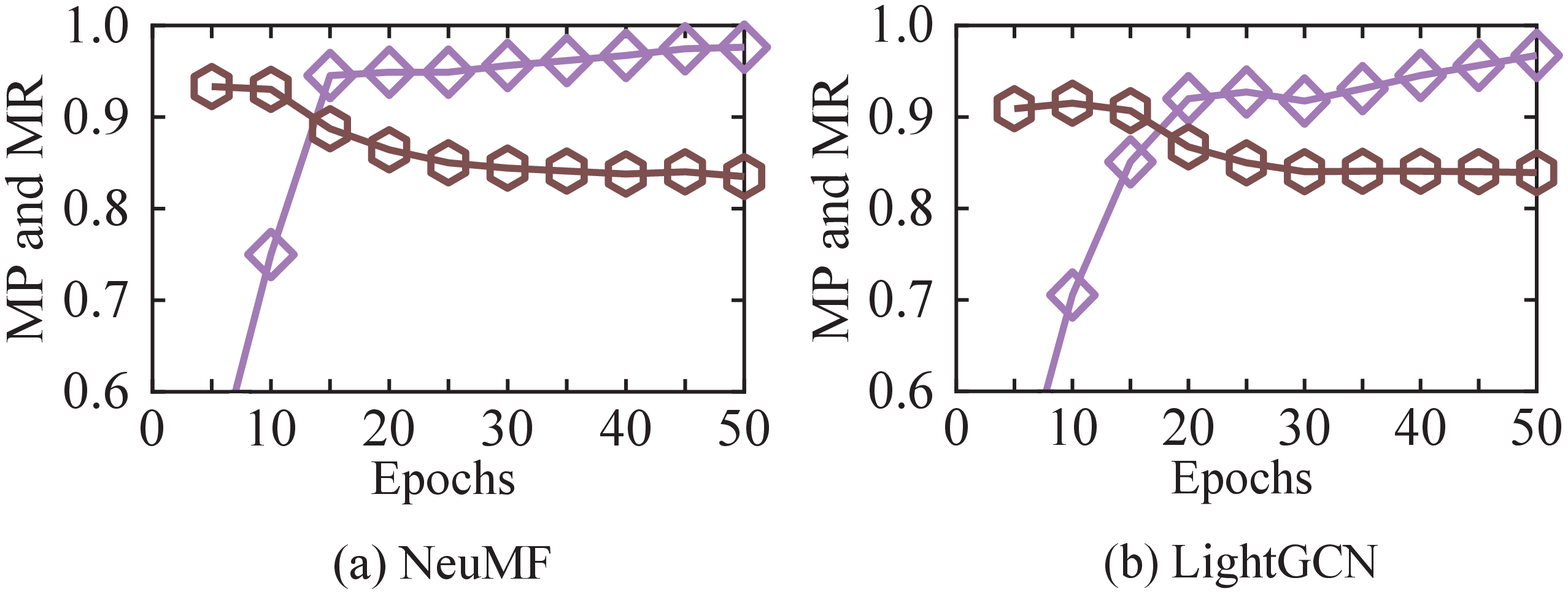}
	\vspace{-15pt}
	\caption{The monotonicity of MP and MR when training NeuMF and LightGCN on the MovieLens dataset.}
	\label{fig:MP-MR}
	\vspace{-10pt}
\end{figure}

\subsubsection{\textbf{Memorization Point Estimation.}} 
\begin{figure}[t]
    \centering
    \includegraphics[width=0.36\textwidth]{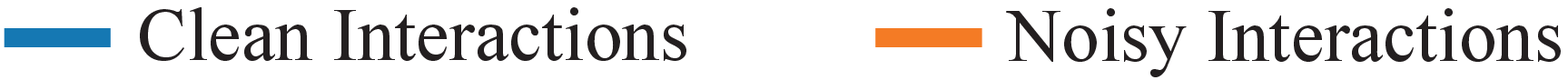}
	\includegraphics[width=0.48\textwidth]{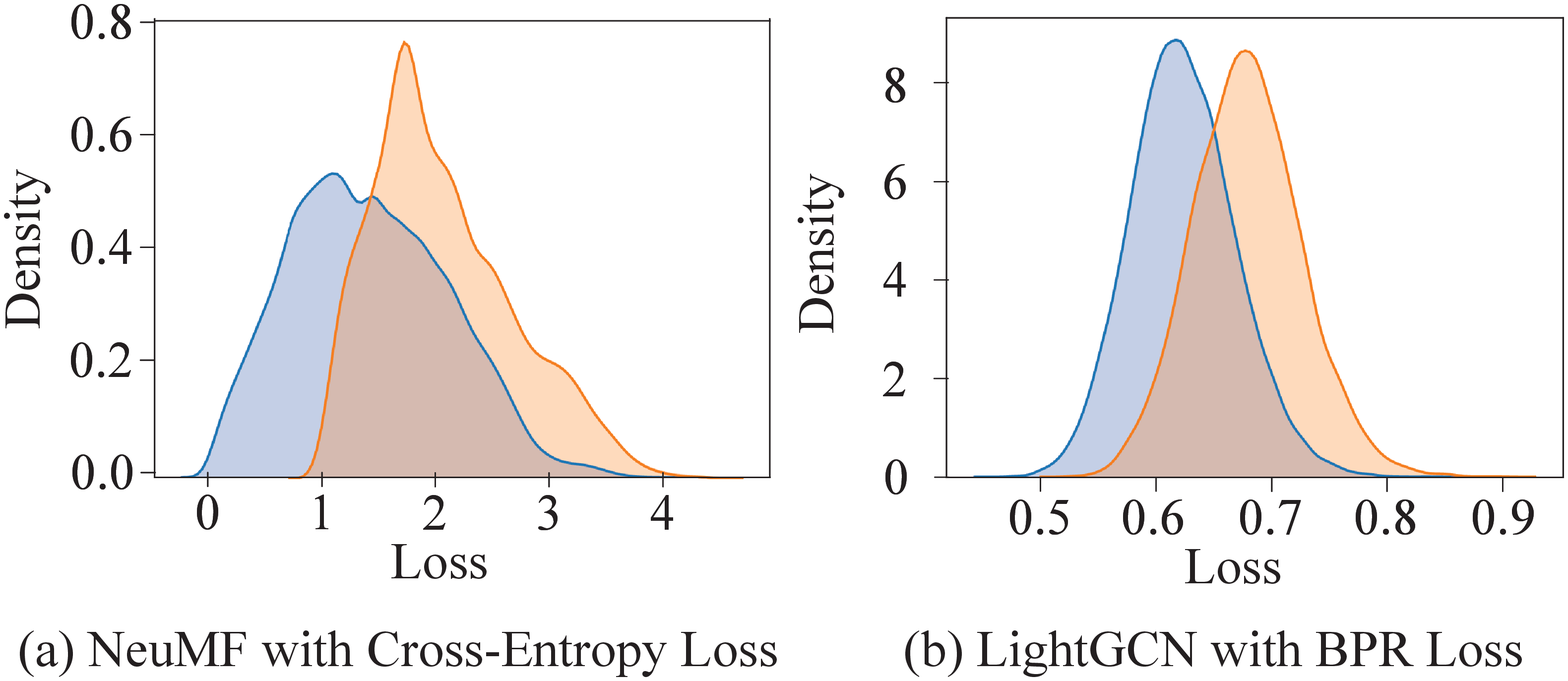}
	\vspace{-15pt}
	\caption{Loss distribution of MovieLens dataset.}
	\label{fig:loss-distribution}
	\vspace{-10pt}
\end{figure}

As shown in Figure~\ref{fig:memorization}, the model predominantly learns clean interactions until the noise-sensitive period begins. In others words, during noise-resistant period, the recommendation model (1) not only learns \textit{sufficient} information from the clean interactions, (2) but also accumulates \textit{some} noise from the noisy implicit feedback. Therefore, we aim to design two metrics to reflect the above two memorization characteristics of the model. Formally, we use $\Set{M}_t$ to denote a set of memorized interactions at epoch $t$, and use $y_{ui}^{*}$ to represent the \textit{true} label of interaction $(u,i)$, which is not available because of the noise in implicit feedback. Inspired by recent advances in robust learning~\cite{nips19meta,kdd21self-trainsition} and recommendation evaluation metrics, we propose two memorization-based metrics, namely, \emph{memorization precision} ($MP$) and \emph{memorization recall} ($MR$), to address the memorization effect of recommendation models respectively:
\begin{gather}\label{equ:MP-MR}
MP_t = \frac{|\Set{R}_t|}{|\Set{M}_t|},  \quad
MR_t = \frac{|\Set{R}_t|}{|\Set{G}|}
\end{gather}where $\Set{R}_t = \{(u,i)\in\Set{M}_t:y_{ui} = y_{ui}^{*}\}$ denotes the set of memorized data whose true labels are consistent with predictions, and $\Set{G} = \{(u,i)\in\Set{D}:y_{ui} = y_{ui}^{*}\}$ is the set of true labeled data in implicit feedback. According to the definition of $MP$ and $MR$, we can conclude that: $MP$ monotonically \textit{decreases} since the model tends to memorize clean data first and then gradually memorizes all the noisy interactions as the training progresses; and $MR$ monotonically \textit{increases} because the model eventually memorizes all clean interactions as the training progresses, as depicted in Figure~\ref{fig:MP-MR} (please refer to Section~\ref{sec:model-analysis} for the theoretical analysis of the monotonicity of the two metrics).
Thus, the best memorization point $t_m$ is naturally the best trade-off epoch when $MP$ and $MR$ share the same value,
\ie $MP_t = MR_t$.
By substituting it into Equation (\ref{equ:MP-MR}), the best memorization point $t_m$ can be calculated as:
\begin{gather}\label{equ:memorization_point}
\Set{M}_{t_m} = |\{(u,i)\in\Set{D}_t:y_{ui} = y_{ui}^{*}\}| = (1-\sigma) |\Set{D}|
\end{gather}
where $\sigma$ is the noise rate of the implicit feedback. Since $\sigma$ is typically unknown, we leverage the difference of loss distributions in clean and noisy data to estimate $\sigma$, as shown in Figure~\ref{fig:loss-distribution}. 

Specifically, we first normalize the loss values of all training interactions and then fit them into a two-component Gaussian Mixture Model (GMM) to model the bi-modal distribution~\cite{icml19loss,iclr20divide_mix} of true- and false-labed samples. The GMM model can be easily trained by using the Expectation-Maximization (EM) algorithm. Hence, we could obtain the probability of an interaction $(u,i)$ being noisy through the posterior probability of loss distributions. Accordingly, the noise rate $\sigma$ is estimated as:
\begin{gather}\label{equ:noise_rate}
\hat{\sigma}  = \Space{E}_{(u,i)\in\Set{D}} [p(\mu|L_{(u,i)}(\theta)]
\end{gather}
where $L_{(u,i)}(\theta)$ is the loss of interaction $(u,i)$ of recommendation model $\theta$, and $\mu$ is the Gaussian component with a larger mean, since noisy data have typically larger loss values. Note that there are many approaches~\cite{pami15noisy,icml19noisy} available to estimate the noise ratio, we choose GMM because it is easy to apply and has stable performance on different noisy datasets.

Therefore, SGDL transits to phase II when the number of memorized interactions reaches
%is greater than or equal to
the estimated clean data size (\ie $|\Set{M}_t| \ge (1-\hat{\sigma})\Set{D}$). Note that, SGDL is able to estimate the memorization point $t_m$, and collect memorized data $\Set{M}_{t_m}$ with nearly \textit{zero} computational overhead and \textit{no} additional supervision.

\subsection{Phase II: Self-Guided Learning}

Phase II aims to robustly learn the recommendation model with the denoising signals of memorized data, which are collected from the training process of Phase I. Previous studies~\cite{wsdm21denoise,arxiv22ensemble} need to pre-specify weighting functions (\eg exponential function), and require additional hyperparameters (\eg threshold) to denoise implicit feedback during training. However, we argue that these methods are fairly hard to be generally applied in real-world recommendation systems due to two issues: (1) The proper weighting schemes heavily rely on the training data, which limits their adaptivity. (2) These methods simply abandon the hard yet clean interactions because the fixed weighting functions fail to distinguish the interactions from noisy data, incurring suboptimal recommendation performance. Thus, we aim to propose a self-guided method that is capable of learning an adaptive weighting function automatically from the collected memorized data to tackle the above two issues.

The above denoising scheme assumes that the memorized data is clean and useful to provide denoising signals for learning user preferences. Nevertheless, as the memorized data is collected by leveraging the memorization effect of recommendation models, it inevitably contains some noise. To prevent the model from being corrupted by such detrimental interactions, we further devise a novel adaptive denoising scheduler to select and use proper memorized data for self-guided learning.

\subsubsection{\textbf{Denoising Learning with Memorized Data.}}
\label{sec:denoising_learning}
During the noise-sensitive period, we aim to enhance the robustness of training by imposing a weight on each sample loss. Here, we consider the memorized data that is able to provide denoising signals to learn the weight for each sample, since it is mostly clean thanks to the memorization effect of the model. Specifically, let $L_k(\theta)$ be the $k$-th sample loss\footnote{The loss can be represented as the pointwise or pairwise loss, according to the optimization target of models. We demonstrate the effectiveness of our method on both kinds of loss functions in Section~\ref{sec:performace}.} of model $\theta$, and $\Set{D}_T$ be the training data. The optimal recommendation model parameter $\theta$ is calculated by minimizing the following weighted loss:
\begin{gather}\label{equ:weighted_loss}
\theta^{*}(\psi) = \argmin_\theta \frac{1}{|\Set{D}_T|} \sum\nolimits_{k}^{|\Set{D}_T|} g(L_k(\theta);\psi)L_k(\theta)
\end{gather}
where $g(L_k(\theta); \psi)$ is the weight on the $k$-th sample loss, and $\psi$ represents the current parameters of weighting function $g(\cdot)$. Here, we formulate $g(\cdot)$ as a simple Multi-Layer Perceptron (MLP) with only one hidden layer, since it is known as a universal approximator for almost any continuous function~\cite{01ann,nips19meta}. To adaptively learn the weight for each sample from memorized data, we optimize the weighting parameters $\psi$ given current optimal $\theta^*(\psi)$:\footnote{Notice that, here $\psi$ is a variable instead of a quantity, which makes $\theta^*(\psi)$ a function of $\psi$, and the gradient in Equation~(\ref{equ:meta_loss}) can be computed.}
\begin{gather}\label{equ:meta_loss}
\psi^* = \argmin_w \frac{1}{|\Set{M}_{t_m}|}\sum\nolimits_m^{|\Set{M}_{t_m}|}L_m(\theta^*(\psi))
\end{gather}
where $L_m(\theta^*(\psi))$ is the $m$-th memorized sample loss given optimal model parameters $\theta^*(\psi)$. Note that the searching for the optimal $\theta^*$ and optimal $\psi^*$ requires two nested loops of optimization (\ie bi-level optimization)~\cite{nips19meta,sigir21autodebias}, which is costly. Hence, we adopt the idea of meta learning~\cite{icml17maml,nips18learing_loss}, and update $\theta$ and $\psi$ alternately in a single loop to guarantee the efficiency of the algorithm. Specifically, as illustrated in Figure~\ref{fig:framework}, we perform the following training procedure in each iteration:
\begin{itemize}[leftmargin=*]
    \item \textbf{Assumed update of $\theta$.} As shown by the blue arrows in Figure~\ref{fig:framework}, we first make an assumed update of $\theta$ with current weight $\psi$:
    \begin{gather}\label{equ:assumed_update_rs}
    \hat{\theta}(\psi)=\theta - \eta_{1}  \frac{1}{|\Set{D}_T|} \sum\nolimits_{k}^{|\Set{D}_T|} g(L_k(\theta);\psi) \nabla_{\theta} L_k(\theta)
    \end{gather}
    where we update $\theta$ using gradient descent with learning rate $\eta_{1}$.
    \item \textbf{Update of $\psi$.} As indicated by the brown arrow in Figure~\ref{fig:framework}, the updates of the weighting parameters $\psi$ can be guided by gradient of the memorized data on the updated model:
    \begin{gather}\label{equ:update_mlp}
    \psi \leftarrow \psi - \eta_{2} \frac{1}{|\Set{M}_{t_m}|}\sum\nolimits_m^{|\Set{M}_{t_m}|} \nabla_{\psi} L_m(\hat{\theta}(\psi))
    \end{gather}
    where $\eta_{2}$ is the learning rate of weighting parameters $\psi$.
    \item \textbf{Actual update of $\theta$.} After receiving the denoising signals (\ie updated parameters $\psi$) from memorized data, we use them to update the model:
    \begin{gather}\label{equ:update_rs}
    \theta \leftarrow \theta- \eta_{1}  \frac{1}{|\Set{D}_T|} \sum\nolimits_{k}^{|\Set{D}_T|} g(L_k(\theta);\psi) \nabla_{\theta} L_k(\theta)
    \end{gather}
\end{itemize}

We use this alternative strategy to optimize the recommendation model and weighting function during the noise-sensitive period, as illustrated in Figure~\ref{fig:framework}. Although this strategy does not guarantee to find the global optimum, it empirically works well in many bi-level optimization problems~\cite{icml17maml,nips19meta,sigir21autodebias}.

\subsubsection{\textbf{Adaptive Denoising Scheduler.}}
\label{sec:adaptive-denoising-scheduler}
Above denoising scheme relies on the memorized data to provide denoising signals for training. However, as memorized data inevitably exhibit noise, integrating the noise of them would degrade the denoising performance.
% Meanwhile, user behaviors are not equally important, some interactions are more likely to reflect the underlying preference of users~\cite{cikm21sequential-denoise,arxiv22ensemble}.

Thus, we propose an adaptive denoising scheduler to select \emph{only} the clean and informative memorized data for denoising learning. Specifically, we define the scheduler as $s$ %\baihua{change the notation as $h$ was used in Section 3.1.1 to refer to the number of most recent memorization histories of (u,i).}\yuntao{Sorry for that, we have changed.}
with parameter $\phi$, and choose two representative factors to quantify the contribution of each memorized data for denoising: (1) the loss $L_m(\theta)$ of $m$-th memorized sample, where $\theta$ is the actual updated parameters of the model; and (2) the gradient similarity of $m$-th memorized sample on assumed updated model parameters $\hat{\theta}$ and actual updated model parameters $\theta$, \ie $\cos\big(\nabla_{\hat{\theta}}L_m(\hat{\theta}), \nabla_{\theta}L_m(\theta)\big)$. Here, we use cosine function as the similarity measurement, and other metrics like inner product can also be applied in practice. The two factors are associated with the learning outcome and learning process of the $m$-th memorized sample, respectively. Specifically, the gradient similarity characterizes the contribution of the memorized sample in the model training. A large loss value may represent a crucial memorized sample if the gradient similarity is also large (\ie the gradient direction of memorized sample is consistent with the optimization of the model); and a large loss value with small gradient similarity may indicate a noisy memorized sample. Considering the two factors simultaneously, we formulate the sampling probability of $m$-th memorized data:
\begin{gather}\label{equ:sampling}
o_m = s\big(L_m(\theta), \cos(\nabla_{\hat{\theta}}L_m(\hat{\theta}), \nabla_{\theta}L_m(\theta)) ;\phi\big) \\
\pi_m = \frac{\exp(o_m;\phi)}{\sum_{i\in\Set{M}_{t_m}} \exp(o_i;\phi)}
\end{gather}
where $o_m$ is the output of the scheduler, and $\pi_m$ is the predicted sampling probability of $m$-th sample. We choose LSTM network~\cite{97lstm} as the scheduler, and feed it with the training factors in each iteration. The intuition behind is that the LSTM could leverage the historical information to capture the prediction variance~\cite{nips17high_variance,yao2021meta}, which shows stable performance in our experiments (different strategies are compared in Section~\ref{sec:ablation-scheduler}). However, it is intractable to directly optimize the scheduler since the sampling process is not differentiable. To make this procedure differentiable and to jointly optimize the scheduler and the model, we apply the Gumbel-Softmax reparameterization trick~\cite{iclr17gumbel} to generate differentiable samples:
\begin{gather}\label{equ:differentiable-sampling}
y_m = \frac{\exp (\log (\pi_m) + \epsilon_m)/\tau}{\sum_{i\in\Set{M}_{t_m}} \exp (\log (\pi_i) + \epsilon_i)/\tau}
\end{gather}
where $\epsilon_m$ is randomly drawn from uniform distribution between 0 and 1, and $\tau$ is the temperature that controls the interpolation between the discrete distribution and continuous categorical densities (we set $\tau=0.05$ for all experiments). Thus, the scheduler is able to decide which memorized data to use according to its contribution to denoising implicit feedback, and adpatively adjust the sampling probability for more informative guided learning.

\subsection{Model Analysis}\label{sec:model-analysis}
\subsubsection{\textbf{Analysis on the monotonicity of $MP$ and $MR$.}} We now prove that $MP$ and $MR$ change monotonically over the training time $t$. Let $\Set{N}_t$ ($= \Set{M}_t \setminus \Set{R}_t$) be the set of memorized data that are falsely predicted by models. Then, $|\Set{N}_{t+1}| / |\Set{N}_{t}| \ge |\Set{R}_{t+1}| / |\Set{R}_t|$ typically holds because the noisy samples are memorized faster than clean samples after the model stabilizes, as depicted in Figure~\ref{fig:MP-MR}. Hence, it is easy to conclude:
{\setlength{\abovedisplayskip}{3pt}
\setlength{\belowdisplayskip}{3pt}
\begin{align}\label{equ:analysis-MP-MR}
|\Set{N}_{t+1}| / |\Set{N}_{t}| &\ge |\Set{R}_{t+1}| / |\Set{R}_t| \\
\implies |\Set{N}_{t+1}||\Set{R}_t| + |\Set{R}_{t+1}| |\Set{R}_t| &\ge |\Set{R}_{t+1}| |\Set{N}_{t}| + |\Set{R}_{t+1}| |\Set{R}_t| \\
\implies (|\Set{N}_{t+1}| + |\Set{R}_{t+1}|)|\Set{R}_{t}| &\ge (|\Set{N}_{t}| + |\Set{R}_{t}|)|\Set{R}_{t+1}| \label{eq:15}
\end{align}}Then, $MP_{t+1}\ge MP_{t}$ can be directly derived from Inequation~(\ref{eq:15}). Besides, considering that the model eventually would memorize all training data~\cite{kdd21self-trainsition}, we assume that $\Set{M}$ would gradually include more observed interactions, including true predicted ones, \ie $|\Set{R}_{t+1}| \ge |\Set{R}_t|$. Consequently, $MR$ increases monotonically since $|\Set{R}_{t+1}|/|\Set{G}|\ge|\Set{R}_{t}|/|\Set{G}|$.

\subsubsection{\textbf{Analysis of the Self-Guide Learning Scheme.}} We then focus on the guidance strategy to explain how memorized data benefit the denoising training. Formally, we follow~\cite{nips19meta}, and utilize chain rule to derive the update function of $\psi$:
\begin{gather}\label{equ:analysis-meta}
\psi \leftarrow \psi + \frac{\eta_1\eta_2}{|\Set{D}_T|}\sum_k^{|\Set{D}_T|} \big(\frac{1}{|\Set{M}_{t_m}|}\sum_m^{|\Set{M}_{t_m}|} G_{mk}\big)\nabla_{\psi}g (L_k(\hat{\theta});\psi)
\end{gather}
where $G_{mk}$ ($ = \nabla_{\theta}L_m(\hat{\theta})^T \nabla_{\theta}L_k(\theta)$) measures the gradient similarity between $m$-th memorized data and $k$-th training data. Thus, for each $k$-th training sample, if its gradient is similar to the average gradient of memorized data, it would be considered as a beneficial one for learning, and its weight tends to be increased. Conversely, the weight of the sample inclines to be suppressed. Therefore, memorized data 
is able to offer a proper weight for each training sample in terms of gradient similarities under the self-guide learning scheme.

\subsubsection{\textbf{Model Size.}} The additional parameters of SGDL come from two parts: (1) the parameters $2d_w$ of weighting function, where $d_w$ is the number of the hidden neurons in one layer MLP; and (2) the $4d_l^2 + 12d_l$ parameters of LSTM unit in the adaptive denoising scheduler, where $d_l$ is the dimension of hidden size. Overall, the additional cost of SDGL is negligible, compared with the tremendous parameters of modern recommendation models.

\subsubsection{\textbf{Time Complexity.}} Assume that the time complexity of the base model is $O(T)$, the additional complexity of SGDL mainly comes from phase II, which consists of two denoising components: (1) the cost of self-guided learning is also $O(T)$, as the alternative optimization scheme takes no more than three times  compared with the normal training; and (2) the computational complexity of denoising adaptive scheduler is $O(|\Set{D}|d_l)$. Therefore, the additional time complexity of SGDL is $O(T + |\Set{D}|d_l)$. Under the same experimental settings (\ie same base model and same embedding size), SGDL achieves better trade-off between efficiency and effectiveness compared with various state-of-the-art models: i) For sample re-weighting and sample selection methods, although most of them are more time-saving than SGDL, they suffer from difficult/expensive hyperparameters tuning and unstable performance, as discussed in Section~\ref{sec:performace}. ii) For robust graph-based methods, SGDL has a complexity that is comparable with them, since graph-based methods typically leverage on extra graph structure to enhance the robustness of the model.

\section{Experiments}
\label{sec:expperiments}
We provide empirical results to demonstrate the effectiveness of our proposed SGDL. The experiments are designed to answer the following research questions:
\begin{itemize}[leftmargin=*]
    \item \textbf{RQ1:} How does SGDL perform, compared with the state-of-the-art denoising methods as well as the state-of-the-art robust recommender methods?
    \item \textbf{RQ2:} How does each component of SGDL
    (\ie memorization point estimation, denoising learning strategy, and adaptive denoising scheduler) affect SGDL?
    \item \textbf{RQ3:} Is SGDL able to distinguish hard yet clean interactions from noisy interactions?
\end{itemize}

\subsection{Experimental Settings}
\label{sec:experimental_settings}

\subsubsection{\textbf{Dataset Description.}}
\label{sec:dataset_description}
We select three real world benchmark datasets to evaluate and compare the performance of SGDL and its competitors. Table~\ref{tab:dataset} lists their statistics.
%recommender performance:
\begin{itemize}[leftmargin=*]
    \item \textbf{Adressa\footnote{\url{https://www.adressa.no/}}} is a news reading dataset from Adressavisen, including user click behaviors and the dwell time for each click. Following previous work~\cite{wsdm21denoise}, clicks with dwell time less than 10 seconds are viewed as noisy interactions.
    \item \textbf{Yelp\footnote{\url{https://www.yelp.com/dataset/challenge}}} is an open recommendation dataset released by the Yelp challenge. We use the 2018 version in our experiments. We follow ~\cite{wsdm21denoise} to mark ratings below 3 as noisy interactions.
    \item \textbf{MovieLens} is a widely used dataset for recommendation, which contains 100,000 movie ratings ranging from 1 to 5. Ratings below 3 are regarded as noisy interactions.
\end{itemize}

\begin{table}[t]
    \centering
    \caption{Statistics of the datasets used in our experiments.}
    \vspace{-10pt}
    \label{tab:dataset}
    \resizebox{0.46\textwidth}{!}{
    \begin{tabular}{c|r|r|r|r}
    \hline
    Dataset          & \#Users & \#Items & \#Interactions & Sparsity \\ \hline\hline
    Adressa      & 212,231  & 6,596  & 419,491      & 99.97\% \\
    MovieLens         & 943  & 1,683  & 100,000      & 93.70\% \\
    Yelp & 45,548 & 57,396  & 1,672,520      & 99.94\% \\ \hline
    \end{tabular}}
    \vspace{-10px}
\end{table}

\begin{table*}[t]
\centering
\vspace{-10pt}
\caption{Performance comparison of different denoising methods on robust recommendation. ``${\dagger}$'' indicates the
improvement of the SGDL over the baseline is significant at the level of 0.05. The highest scores are in Bold. R and N refer to Recall and NDCG, respectively.  
}
\vspace{-10pt}
\label{tab:overall-performance}
\resizebox{0.98\textwidth}{!}{
\begin{tabular}{c|c|c|c|c|c|c|c|c|c|c|c|c|c}
\hline
\multicolumn{2}{c|}{\textbf{Database}}                                         & \multicolumn{4}{c|}{\textbf{Adressa}}                                    & \multicolumn{4}{c|}{\textbf{MovieLens}}                                    & \multicolumn{4}{c}{\textbf{Yelp}}                                     \\ \hline
\multicolumn{1}{c|}{\textbf{Base Model}} & \multicolumn{1}{c|}{\textbf{Method}} & \multicolumn{1}{c|}{\textbf{R@5}} & \multicolumn{1}{c|}{\textbf{R@20}} &
\multicolumn{1}{c|}{\textbf{N@5}} & \multicolumn{1}{c|}{\textbf{N@20}} & \multicolumn{1}{c|}{\textbf{R@5}} & \multicolumn{1}{c|}{\textbf{R@20}} &
\multicolumn{1}{c|}{\textbf{N@5}} & \multicolumn{1}{c|}{\textbf{N@20}} & \multicolumn{1}{c|}{\textbf{R@5}} & \multicolumn{1}{c|}{\textbf{R@20}} &
\multicolumn{1}{c|}{\textbf{N@5}} & \multicolumn{1}{c}{\textbf{N@20}} \\ \hline
\multirow{6}{*}{NeuMF}                & Normal   &0.1533$^{\dagger}$	&0.3208$^{\dagger}$	&0.1224$^{\dagger}$	&0.1808$^{\dagger}$	&0.1023$^{\dagger}$	&0.2687$^{\dagger}$	&0.2890$^{\dagger}$	&0.2765$^{\dagger}$	&0.0129$^{\dagger}$	&0.0393$^{\dagger}$	&0.0129$^{\dagger}$	&0.0215$^{\dagger}$	   \\
                                      & WBPR     &0.1538$^{\dagger}$	&0.3207$^{\dagger}$	&0.1225$^{\dagger}$	&0.1809$^{\dagger}$	&0.1025$^{\dagger}$	&0.2689$^{\dagger}$	&0.2891$^{\dagger}$	&0.2769$^{\dagger}$	&0.0128$^{\dagger}$	&0.0392$^{\dagger}$	&0.0127$^{\dagger}$	&0.0214$^{\dagger}$	  \\
                                      & IR       &0.1541$^{\dagger}$	&0.3212$^{\dagger}$	&0.1229$^{\dagger}$	&0.1830$^{\dagger}$	&0.1054$^{\dagger}$	&0.2704$^{\dagger}$	&0.2928$^{\dagger}$	&0.2758$^{\dagger}$	&0.0132$^{\dagger}$	&0.0407$^{\dagger}$	&0.0131$^{\dagger}$	&0.0229$^{\dagger}$	   \\
                                      & T-CE     &0.1537$^{\dagger}$	&0.3220$^{\dagger}$	&0.1267$^{\dagger}$	&0.1839$^{\dagger}$	&0.1025$^{\dagger}$	&0.2821$^{\dagger}$	&0.2923$^{\dagger}$	&0.2845$^{\dagger}$	&0.0119$^{\dagger}$	&0.0396$^{\dagger}$	&0.0119$^{\dagger}$	&0.0211$^{\dagger}$	   \\
                                      & DeCA     &0.1597	&0.3205$^{\dagger}$	&0.1226$^{\dagger}$	&0.1799$^{\dagger}$	&0.1024$^{\dagger}$	&0.2723$^{\dagger}$	&0.2904$^{\dagger}$	&0.2801$^{\dagger}$	&0.0129$^{\dagger}$	&0.0394$^{\dagger}$	&0.0129$^{\dagger}$	&0.0216$^{\dagger}$	   \\  \cline{2-14}
                                      & SGDL     &\textbf{0.1598}	&\textbf{0.3291}	&\textbf{0.1272}	&\textbf{0.1853}	&\textbf{0.1135}	&\textbf{0.2844}	&\textbf{0.3279}	&\textbf{0.3032}	&\textbf{0.0155}	&\textbf{0.0469}	&\textbf{0.0158}	&\textbf{0.0260}	   \\ \hline \hline
\multirow{6}{*}{CDAE}                 & Normal   &0.1445$^{\dagger}$	&0.3159$^{\dagger}$	&0.0987$^{\dagger}$	&0.1886$^{\dagger}$	&0.0904$^{\dagger}$	&0.2185$^{\dagger}$	&0.2617$^{\dagger}$	&0.2356$^{\dagger}$	&0.0145$^{\dagger}$	&0.0436$^{\dagger}$	&0.0149$^{\dagger}$	&0.0277$^{\dagger}$	   \\
                                      & WBPR     &0.1443$^{\dagger}$	&0.3158$^{\dagger}$	&0.0987$^{\dagger}$	&0.1890$^{\dagger}$	&0.0908$^{\dagger}$	&0.2184$^{\dagger}$	&0.2619$^{\dagger}$	&0.2346$^{\dagger}$	&0.0148$^{\dagger}$	&0.0437$^{\dagger}$	&0.0151$^{\dagger}$	&0.0278$^{\dagger}$	   \\
                                      & IR       &0.1444	&0.3152$^{\dagger}$	&0.0981$^{\dagger}$	&0.1893$^{\dagger}$	&0.0909$^{\dagger}$	&0.2186$^{\dagger}$	&0.2612$^{\dagger}$	&0.2358$^{\dagger}$	&0.0153$^{\dagger}$	&0.0438	&0.0152$^{\dagger}$	&0.0278$^{\dagger}$	  \\
                                      & T-CE     &0.1415$^{\dagger}$	&0.3106$^{\dagger}$	&0.0991	&0.1840$^{\dagger}$	&0.0912$^{\dagger}$	&0.2158$^{\dagger}$	&0.2642	&0.2386$^{\dagger}$	&0.0147$^{\dagger}$	&\textbf{0.0439}	&0.0151$^{\dagger}$	&0.0279$^{\dagger}$	  \\
                                      & DeCA     &0.1447$^{\dagger}$	&0.3159$^{\dagger}$	&0.0991	&0.1888$^{\dagger}$	&0.0917$^{\dagger}$	&0.2189$^{\dagger}$	&0.2641	&0.2378$^{\dagger}$	&0.0158$^{\dagger}$	&0.0438	&0.0154$^{\dagger}$	&0.0292$^{\dagger}$	   \\ \cline{2-14}
                                      & SGDL     &\textbf{0.1450}	&\textbf{0.3181}	&\textbf{0.0993}	&\textbf{0.1956}	&\textbf{0.0921}	&\textbf{0.2220}	&\textbf{0.2643}	&\textbf{0.2404}	&\textbf{0.0162}	&\textbf{0.0439}	&\textbf{0.0172}	&\textbf{0.0296}	   \\ \hline \hline
\multirow{7}{*}{NGCF}                 & Normal   &0.0769$^{\dagger}$	&0.1322$^{\dagger}$	&0.0571$^{\dagger}$	&0.0769$^{\dagger}$	&0.1285$^{\dagger}$	&0.3103$^{\dagger}$	&0.3694$^{\dagger}$	&0.3392$^{\dagger}$	&0.0267$^{\dagger}$	&0.0736$^{\dagger}$	&0.0262$^{\dagger}$	&0.0417$^{\dagger}$	  \\
                                      & WBPR     &0.0770$^{\dagger}$	&0.1324$^{\dagger}$	&0.0572$^{\dagger}$	&0.0769$^{\dagger}$	&0.1287$^{\dagger}$	&0.3105$^{\dagger}$	&0.3692$^{\dagger}$	&0.3395$^{\dagger}$	&0.0265$^{\dagger}$	&0.0739$^{\dagger}$	&0.0265$^{\dagger}$	&0.0417$^{\dagger}$	   \\
                                      & IR       &0.0772$^{\dagger}$	&0.1337$^{\dagger}$	&0.0570$^{\dagger}$	&0.0768$^{\dagger}$	&0.1280$^{\dagger}$	&0.3104$^{\dagger}$	&0.3701$^{\dagger}$	&0.3395$^{\dagger}$	&0.0269$^{\dagger}$	&0.0737$^{\dagger}$	&0.0261$^{\dagger}$	&0.0412$^{\dagger}$	   \\
                                      & DeCA     &0.0760$^{\dagger}$	&0.1326$^{\dagger}$	&0.0571$^{\dagger}$	&0.0766$^{\dagger}$	&0.1304$^{\dagger}$	&0.3113$^{\dagger}$	&0.3729$^{\dagger}$	&0.3401	&0.0277	&0.0739$^{\dagger}$	&0.0262	&0.0418	   \\ \cline{2-14}
                                      & SGCN     &0.0773$^{\dagger}$	&0.1336$^{\dagger}$	&0.0543$^{\dagger}$	&0.0770	&0.1288$^{\dagger}$	&0.3112$^{\dagger}$	&\textbf{0.3768}	&0.3401	&0.0267$^{\dagger}$	&0.0734$^{\dagger}$	&0.0265	&\textbf{0.0443} \\
                                      & SGL      &0.0775$^{\dagger}$	&0.1345	&0.0576	&0.0768$^{\dagger}$	&0.1303$^{\dagger}$	&0.3141$^{\dagger}$	&0.3763$^{\dagger}$	&0.3360$^{\dagger}$	&\textbf{0.0279}	&\textbf{0.0750}	&0.0264$^{\dagger}$	&0.0409$^{\dagger}$	\\ \cline{2-14}
                                      & SGDL     &\textbf{0.0788}	&\textbf{0.1347}	&\textbf{0.0579}	&\textbf{0.0771}	&\textbf{0.1309}	&\textbf{0.3186}	&0.3745	&\textbf{0.3404}	&0.0273	&0.0746	&\textbf{0.0267}	&0.0420	   \\ \hline \hline
\multirow{7}{*}{LightGCN}             & Normal   &0.0951$^{\dagger}$	&0.1817$^{\dagger}$	&0.0713$^{\dagger}$	&0.0994$^{\dagger}$	&0.1258$^{\dagger}$	&0.3173$^{\dagger}$	&0.3678$^{\dagger}$	&0.3358$^{\dagger}$	&0.0334$^{\dagger}$	&0.0912$^{\dagger}$	&0.0332$^{\dagger}$	&0.0515$^{\dagger}$	   \\
                                      & WBPR     &0.0958$^{\dagger}$	&0.1845$^{\dagger}$	&0.0733$^{\dagger}$	&0.1006$^{\dagger}$	&0.1262$^{\dagger}$	&0.3189$^{\dagger}$	&0.3701$^{\dagger}$	&0.3510	&0.0333$^{\dagger}$	&0.0911$^{\dagger}$	&0.0331$^{\dagger}$	&0.0512$^{\dagger}$	   \\
                                      & IR       &0.0953$^{\dagger}$	&0.1822$^{\dagger}$	&0.0726$^{\dagger}$	&0.1003$^{\dagger}$	&0.1285$^{\dagger}$	&0.3194$^{\dagger}$	&0.3681$^{\dagger}$	&0.3361$^{\dagger}$	&0.0305$^{\dagger}$	&0.0909$^{\dagger}$	&0.0326$^{\dagger}$	&0.0510$^{\dagger}$	   \\
                                      & DeCA     &0.0974$^{\dagger}$	&0.1855$^{\dagger}$	&0.0758$^{\dagger}$	&0.1162$^{\dagger}$	&0.1293$^{\dagger}$	&0.3076$^{\dagger}$	&0.3575$^{\dagger}$	&0.3270$^{\dagger}$	&0.0337	&0.0911$^{\dagger}$	&0.0332$^{\dagger}$	&0.0524	   \\ \cline{2-14}
                                      & SGCN     &0.0941$^{\dagger}$	&0.1899$^{\dagger}$	&0.0765$^{\dagger}$	&0.1160$^{\dagger}$	&0.1282$^{\dagger}$	&0.3210$^{\dagger}$	&0.3602$^{\dagger}$	&0.3318$^{\dagger}$	&0.0335$^{\dagger}$	&0.0916	&\textbf{0.0346}	&\textbf{0.0528} \\
                                      & SGL      &0.0980$^{\dagger}$	&0.1770$^{\dagger}$	&0.0741$^{\dagger}$	&0.0999$^{\dagger}$	&0.1299$^{\dagger}$	&0.3156$^{\dagger}$	&0.3638$^{\dagger}$	&0.3343$^{\dagger}$	&\textbf{0.0341}	&0.0915	&0.0344	&0.0526	   \\ \cline{2-14}
                                      & SGDL     &\textbf{0.1134}	&\textbf{0.2105}	&\textbf{0.0844}	&\textbf{0.1178}	&\textbf{0.1378}	&\textbf{0.3335}	&\textbf{0.3844}	&\textbf{0.3513}	&0.0339	&\textbf{0.0918}	&0.0341	&0.0525	   \\ \hline
\end{tabular}}
\vspace{-10pt}
\end{table*}

\subsubsection{\textbf{Evaluation Metrics.}} We adopt cross-validation to verify the performance. Specifically, we follow~\cite{wsdm21denoise,arxiv22ensemble} to split the interactions into the training set, validation set, and clean test set with the ratio of 8:1:1. The performance is measured by two widely used valuation protocols~\cite{sigir19ngcf,sigir20lighGCN}: Recall@$K$ and NDCG@$K$, where $K$ is set as 5 and 20 by default. We report the average metrics for all users in the test set. 

\subsubsection{\textbf{Baselines.}} We select four state-of-the-art recommendation methods as the base model $f$ of SGDL:
\begin{itemize}[leftmargin=*]
    \item \textbf{NeuMF}~\cite{www17nfm} is a state-of-the-art model, which generalizes Factorization Machines (FM) with a Multi-Layer Perceptron (MLP).
    \item \textbf{CDAE}~\cite{wsdm16cdae} is denoising auto-encoder model, which corrupts the interactions with random noises, and then employs a MLP model to reconstruct the original input.
    \item \textbf{NGCF}~\cite{sigir19ngcf} is a graph model, which applies graph convolution network (GCN) to encode high-order collaborative signals in user-item bipartite graph.
    \item \textbf{LightGCN}~\cite{sigir20lighGCN} is a state-of-the-art graph model, which simplifies the design of GCN by discarding the nonlinear feature transformations for recommendation.
\end{itemize}

We train the base models with different ranking loss functions to demonstrate the universality of SGDL. Specifically, we train NeuMF and CDAE with binary cross-entropy (BCE) loss, and train NGCF and LightGCN with BPR loss~\cite{uai09BPRloss}. Each model is trained with the following denoising approaches:
\begin{itemize}[leftmargin=*]
    \item \textbf{Normal} is trained with the original architecture design, without any denoising consideration.
    \item \textbf{WBPR}~\cite{kdd12ranking} is a sample selection method, which considers the popular but uninteracted items are likely to be real negative ones.
    \item \textbf{IR}~\cite{mm21occf} is the state-of-the-art sample selection method, which interactively relabels uncertain samples to mitigate the noise in both observed and unobserved interactions.
    \item \textbf{T-CE}~\cite{wsdm21denoise} is the state-of-the-art sample re-weighting method, which uses the Truncated BCE to assign zero weights to large-loss examples with a dynamic threshold. Note that, this denoising approach can only be used for BCE loss, and thus, we implement it with NeuMF and CDAE for comparison.
    \item \textbf{DeCA}~\cite{arxiv22ensemble} is a newly proposed sample re-weighting method, which considers the disagreement predictions of noisy samples across different models, and minimizes KL-divergence between the two models’ predictions to enhance the robustness of models.
\end{itemize}

In addition, we also compare SGDL with the state-of-the-art robust graph-based methods to further confirm the effectiveness of our model. Note that the methods can only be applied to graph-based recommenders (\ie NGCF and LightGCN in our experiments), since they regard the noisy interactions as noisy edges, and devise enhanced graph learning methods for robust recommendation.  %\baihua{Why SGCN/SGL are only applied to certain baseline models but not all? In addition, can we group denosing approaches together and graph robust recommender models together in Table 2 to provide a clearer view (say adding two hozirontal lines to further partition the results)? }\yuntao{Because SGCN/SGL is orignally designed for graph-based RS, they cannot apply to NeuMF and CDAE because of the different network architecture. We add them for comparison because they are the most noticeable papers in SIGIR21, as well as to show our effectiveness. And sure, I will check and modify the table now.}\baihua{No harm to include one sentence to highlight the following two models can only be applied to the last two baselines but not all.}
\begin{itemize}[leftmargin=*]
    \item \textbf{SGCN}~\cite{sigir21mask} is the state-of-the-art graph structure enhanced method, which attaches the GCN layers with a trainable stochastic binary mask to prune noisy edges in user-item bipartite graph.
    \item \textbf{SGL}~\cite{sigir21sgl} is the state-of-the-art self-supervised graph method, which designs different graph views to mine hard negatives and denoise noise in implicit feedback. We choose the Edge Dropout (ED) view as auxiliary supervision signal as it performs the best for most datasets.
\end{itemize}

\subsubsection{\textbf{Parameter Settings.}}
We implement SGDL in Pytorch, and have released our implementations\footnote{\url{https://github.com/ZJU-DAILY/SGDL}} (codes, parameter settings, andtraining logs) to facilitate reproducibility. We use recommended parameter settings for all models, and optimize them with Adam~\cite{iclr2015adam} optimizer. The Xavier initializer~\cite{aistats10Xavier} is used to initialize the model parameters. We set the batch size as 128 for MovieLens, 1024 for Adressa, and 2048 for Yelp due to the different sizes of each dataset. The learning rate of CDAE, NeuMF, and LightGCN is tuned as 0.001; and for NGCF, the learning rate is set as 0.0001. For NeuMF, the embedding size is 32, and the number of layers is 3. For CDAE, the hidden size is 100, and the dropout ratio is set to 0.5. For LightGCN, we set the embedding size to 64, the number of layers to 3, and train it without dropout. For NGCF, the embedding size and layers are the same as LightGCN; and the node dropout rate is set to 0.1. For SGDL, we set $\eta_1 = \eta_2$ for denoising learning, and tune the learning rate amongst $\{10^{-4},10^{-3},10^{-2}\}$ for two phases, respectively. The hidden size of MLP and LSTM unit is 64; the length of memorization history $h$ is tuned among $\{2, 5, 10, 20\}$. Note that, the hyperparameters of base models keep exactly the same across all training methods for a fair comparison.

\subsection{Performance Comparison (RQ1)}
\label{sec:performace}
We begin with the performance comparison \wrt recall@$K$ and NDCG@$K$, where we test two values (5 and 20) of $K$.
The experimental results are reported in Table~\ref{tab:overall-performance}, and we find that:

\begin{itemize}[leftmargin=*]
    \item The proposed SGDL can effectively improve the performance of all base models, and outperform all denoising methods over three datasets. Besides, even if the base model is designed to be robust against noisy implicit feedback (\ie CDAE), our method can still boost its performance by a large margin. We attribute these improvement to the memorization-based denoising schemes of SGDL: (1) By tracing the memorization states of data in the noise-resistant period, SGDL is able to collect memorized interactions during training process to provide valuable denoising signals without any supervision. In contrast, none of the baselines considers explicitly characterizing data from the memorization perspective. (2) Benefiting from our learning to weight strategy and adaptive denoising scheduler, SGDL can adaptively select clean and informative samples from memorized interactions, and use them to guide the learning process of the model. However, other re-weighting baselines (\eg T-CE and DeCA) are insufficient to provide proper weight for each interaction since they do not have memorized data as guidance.
    \item Jointly analyzing the performance of SGDL across three datasets, we find that the improvement on MovieLens dataset are less significant than that on other datasets. One possible reason is that the sparsity of MovieLens dataset is denser than the sparsity of Yelp and Adressa. Accordingly, there are sufficient interactions to identify user behavior patterns, which offsets the impact of noisy implicit feedback.
    \item Jointly analyzing the performance of SGDL across the recommenders, we observe that the relative improvement (\ie the performance of SGDL over the strongest baselines) on NeuMF and CDAE are more substantial than that on graph-based methods (\ie NGCF and LightGCN). This is because our method is model-agnostic, which does not take the graph structure into consideration. In comparison, both SGCN and SGL are designed for graph-based recommendations, and thus, they can leverage the graph structure information to yield better performance. Nonetheless, our method still achieves the best results on most cases, which demonstrate the superior robustness of SGDL.
    \item All the denoising approaches show better results compared with normal training in most cases, which indicate the necessities of denoising implicit feedback for recommendations. The results are consistent with prior studies~\cite{wsdm21denoise,arxiv22ensemble}.
    \item The performance of sample selection methods (\ie WBPR and IR) is rather unstable across different recommenders, compared with sample re-weighting methods.
    This is reasonable since sample selection methods highly depend on the sampling distribution, which makes their performance unstable. For instance, although IR achieves some improvement on Yelp dataset, it performs worse than the base model (\ie CDAE) on Adressa dataset.
    \item Robust Graph-based methods (\ie SGCN and SGL) achieve competitive or even the best performance against other methods. We attribute such improvement to their carefully designed graph structures, which is able to prune noisy and insignificant edges for clean information propagation. However, the methods can only be applied to graph-based models, while SGDL is able to be easily integrated with any learning-based recommender systems, and achieves comparable or even better performance.
\end{itemize}

\begin{table*}[t]
\centering
\vspace{-10pt}
\caption{Impact of denoising Learning and adaptive denoising scheduler.}
\vspace{-10pt}
\label{tab:impact-of-DLS-ADS}
\resizebox{0.99\textwidth}{!}{
\begin{tabular}{c|c|c|c|c|c|c|c|c|c|c|c|c|c}
\hline
\multicolumn{2}{c|}{\textbf{Database}}                                         & \multicolumn{4}{c|}{\textbf{Adressa}}                                    & \multicolumn{4}{c|}{\textbf{MovieLens}}                                    & \multicolumn{4}{c}{\textbf{Yelp}}                                     \\ \hline
\multicolumn{1}{c|}{\textbf{Base Model}} & \multicolumn{1}{c|}{\textbf{Method}} & \multicolumn{1}{c|}{\textbf{R@5}} & \multicolumn{1}{c|}{\textbf{R@20}} &
\multicolumn{1}{c|}{\textbf{N@5}} & \multicolumn{1}{c|}{\textbf{N@20}} & \multicolumn{1}{c|}{\textbf{R@5}} & \multicolumn{1}{c|}{\textbf{R@20}} &
\multicolumn{1}{c|}{\textbf{N@5}} & \multicolumn{1}{c|}{\textbf{N@20}} & \multicolumn{1}{c|}{\textbf{R@5}} & \multicolumn{1}{c|}{\textbf{R@20}} &
\multicolumn{1}{c|}{\textbf{N@5}} & \multicolumn{1}{c}{\textbf{N@20}} \\ \hline
\multirow{2}{*}{NeuMF}                & w/o DLS   & 0.1528	&0.3107	&0.1211	&0.1794	&0.1055	&0.2690	&0.2911	&0.2774	&0.0136	&0.0397	&0.0131	&0.0218   \\
                                      & w/o ADS   & 0.1576	&0.3285	&0.1255	&0.1801	&0.1097	&0.2801	&0.3210	&0.3008	&0.0146	&0.0438	&0.0146	&0.0259  \\ \hline
\multirow{2}{*}{LightGCN}             & w/o DLS   & 0.0964	&0.1810	&0.0702	&0.0985	&0.1244	&0.3159	&0.3688	&0.3349	&0.0330	&0.0909	&0.0331	&0.0513   \\
                                      & w/o ADS   & 0.1013	&0.1995	&0.0811	&0.1007	&0.1316	&0.3328	&0.3824	&0.3502	&0.0338	&0.0914	&0.0340	&0.0521 \\ \hline
\end{tabular}}
\vspace{-5pt}
\end{table*}

\subsection{Study of SDGL (RQ2)}
\label{sec:ablation_study}
As the memorization and self-guided denoising learning are the core of SGDL, we also conduct ablation studies to investigate their effectiveness. Specifically, how the presence of denoising learning and adaptive denoising scheduler, the estimation of memorization point, and the design of scheduler affect our model.

\subsubsection{\textbf{Impact of Denoising Learning \& Scheduler.}}
\label{sec:ablation-weight}

We first evaluate the effectiveness of the denoising learning scheme and adaptive denoising scheduler. To this end, two variants of SGDL are constructed by (1) discarding the denoising learning strategy in noise-sensitive period, called SGDL$_{\text{w/o DLS}}$; and (2) removing the scheduler and directly using all memorized data for denoising learning, named SGDL$_{\text{w/o ADS}}$. We summarize the results of NeuMF and LightGCN in Table~\ref{tab:impact-of-DLS-ADS}, while skip the results of other models (\ie CDAE and NGCF) because they demonstrate similar trends, and we have space limitation.

Obviously, compared with SGDL in Table~\ref{tab:overall-performance}, removing the denoising learning scheme (\ie SGDL$_{\text{w/o DLS}}$) dramatically reduces the predictive accuracy, indicating the necessity of self-guided learning. To be more
specific, SGDL$_{\text{w/o DLS}}$ only trains with memorized data in the noise-sensitive period, and thus, it is insufficient to learn user true preference with hard yet clean interactions. Besides, directly leveraging all memorized data as denoising signals would inevitably introduce some noise, and hence, the SGDL$_{\text{w/o ADS}}$ also underperforms the complete model.

\subsubsection{\textbf{Estimation of Memorization Point.}}
\label{sec:ablation-mpoint}

We then verify the estimation of memorization point, since it plays an important role in our model to transit from noise-resistant period to noise-sensitive period. Specifically, we explore the performance change of SGDL by increasing or decreasing the estimated noisy ratio $\hat{\sigma}$ to force early or late transition. The results of SGDL on NeuMF and LightGCN are presented in Table~\ref{tab:impact-of-memoirzation-point}. We observe that:
\begin{itemize}[leftmargin=*]
    \item Generally speaking, the best performance is achieved at the estimated memorization point. When the memorization point more deviates from the estimated one, the performance tends to be worse. This means the existence of the best memorization point and the effectiveness of our estimation.
    \item It is worth mentioning that, when the memorization point is slightly earlier than the estimated value, it also shows good performance. We attribute this to the clean memorized data: when we transit the model to noise-sensitive period earlier, it is more likely that most of the memorized interactions are clean, which benefits the following self-guided learning. On the contrary, when we delay the memorization point, the memorized data tends to contain more noisy samples, incurring suboptimal performance.
\end{itemize}

\begin{table}[t]
\centering
\caption{Impact of estimation of memorization point. R@20 is used to evaluate the performance, and Est. denotes estimated memorization point.}
\vspace{-10pt}
\label{tab:impact-of-memoirzation-point}
\resizebox{0.47\textwidth}{!}{
\begin{tabular}{c|c|c|c|c|c|c}
\hline
\multicolumn{2}{c|}{\textbf{Memorization Point}} & \multicolumn{2}{c|}{\textbf{Early}} & \multicolumn{1}{c|}{\textbf{Est.}}  &\multicolumn{2}{c}{\textbf{Late}}   \\ \hline
\multicolumn{1}{c|}{\textbf{Base Model}} & \multicolumn{1}{c|}{\textbf{Database}} & \multicolumn{1}{c|}{+10\%} & \multicolumn{1}{c|}{+5\%} &
\multicolumn{1}{c|}{+0\%} & \multicolumn{1}{c|}{-5\%} & \multicolumn{1}{c}{-10\%} \\ \hline
\multirow{3}{*}{NeuMF}                & Adressa     &0.3221	&0.3275	&\textbf{0.3291}	&0.3256	&0.3203  \\
                                      & MovieLens   &0.2810	&\textbf{0.2851}	&0.2844	&0.2757	&0.2704	 \\
                                      & Yelp        &0.458	&0.4420	&\textbf{0.0469}	&0.4430	&0.4370	 \\ \hline
\multirow{3}{*}{LightGCN}             & Adressa     &0.2006	&\textbf{0.2114}	&0.2105	&0.2017	&0.1990 \\
                                      & MovieLens   &0.3321	&0.3325	&\textbf{0.3335}	&0.3262	&0.3198 \\
                                      & Yelp        &0.0895	&0.0912	&\textbf{0.0918}	&0.0904	&0.0887   \\ \hline
\end{tabular}}
\vspace{-10pt}
\end{table}

\subsubsection{\textbf{The Design of Adaptive Denoising Scheduler.}}
\label{sec:ablation-scheduler}
We also investigate the design of adaptive denoising scheduler. Specifically, we propose three different approaches to evaluate the denoising contributions of each memorized data: i) Rank memorized data according to their sums of two factors (\ie normalized gradient similarity and loss value), and simply pick top-$F$ memorized data as informative denoising signals. To keep the picked data clean, we set $F$ as the half size of memorized data. ii) Choose a one-layer MLP as the scheduler, and train the scheduler with the strategy presented in Section~\ref{sec:adaptive-denoising-scheduler}. iii) Select the LSTM as the scheduler, and leverage the historical factors to predict the sampling probabilities. We report the performance of three sampling approaches with Recall@20 and NDCG@20 in Table~\ref{tab:impact-of-ADS}, while Recall@5 and NDCG@5 are omitted due to limited space. We have the following observations:
\begin{itemize}[leftmargin=*]
    \item  By simply choosing top-$F$ memorized data with the highest sums as denoising signals, the performance drops and becomes even worse than the normal training. We attribute such degradation to the non-linear correlation of two factors: the large loss value may not necessarily mean the memorized sample is beneficial to denoising, as it can also be noisy if the gradient similarity is small. Directly summing up the two factors is unable to properly capture the characteristic of memorized data, and thus fails to provide reliable denoising signals for self-guided learning.
    \item The scheduler with LSTM consistently achieves the best performance cross the datasets. This is because LSTM can leverage the historical factors information to predict more accurate and stable sampling probability for memorized data. The results are consistent with previous robust learning studies~\cite{nips17high_variance,yao2021meta}.
\end{itemize}

\begin{table}[t]
\centering
\caption{Impact of the design of adaptive denoising scheduler.}
\vspace{-10pt}
\label{tab:impact-of-ADS}
\resizebox{0.48\textwidth}{!}{
\begin{tabular}{c|c|c|c|c|c|c|c}
\hline
\multicolumn{2}{c|}{\textbf{Database}}                                         & \multicolumn{2}{c|}{\textbf{Yelp}}                                    & \multicolumn{2}{c|}{\textbf{MovieLens}}                                    & \multicolumn{2}{c}{\textbf{Adressa}}                                     \\ \hline
\multicolumn{1}{c|}{\textbf{Base Model}} & \multicolumn{1}{c|}{\textbf{ADS}}  & \multicolumn{1}{c|}{\textbf{R@20}} &
\multicolumn{1}{c|}{\textbf{N@20}}  & \multicolumn{1}{c|}{\textbf{R@20}}
 & \multicolumn{1}{c|}{\textbf{N@20}}  & \multicolumn{1}{c|}{\textbf{R@20}} &\multicolumn{1}{c}{\textbf{N@20}} \\ \hline
\multirow{3}{*}{NeuMF}                & top-$F$   &0.3106	&0.1766	&0.2590	&0.2705	&0.0386	&0.0212		   \\
                                      & MLP   &0.3257	&0.1842	&0.2749	&0.2841	&0.0392	&0.0225			  \\
                                      & LSTM   &0.3291	&0.1853	&0.2844	&0.3032	&0.0469	&0.0260			  \\\hline
\multirow{3}{*}{LightGCN}             & top-$F$   &0.1810	&0.0985	&0.3159	&0.3349	&0.0909	&0.0513			   \\
                                      & MLP   &0.2066	&0.0103	&0.3284	&0.3427	&0.0912	&0.0512		  \\
                                      & LSTM   &0.2105	&0.1178	&0.3335	&0.3513	&0.0918	&0.0525		 \\ \hline
\end{tabular}}
\vspace{-10pt}
\end{table}

\begin{figure}[t]
    \vspace{10pt}
    \centering
	\includegraphics[width=0.48\textwidth]{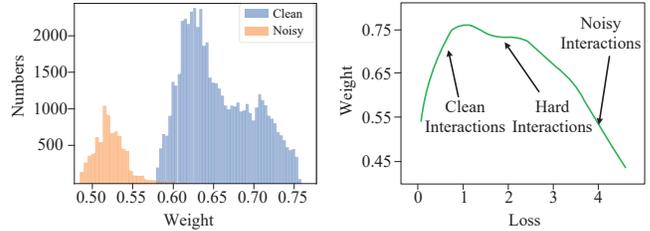}
	\vspace{-20pt}
	\caption{(a) Sample weight distribution on MovieLens dataset. (b) Learned weighting function on MovieLens dataset.}
	\label{fig:visualization}
	\vspace{-10pt}
\end{figure}

\vspace{3pt}
\subsection{Learned Weights of SGDL (RQ3)}
\label{sec:method-interpretation}
In this section, we visualize the learned weights of SGDL to offer an intuitive impression of our denoising performance. Specifically, we train NeuMF with SGDL on the MovieLens dataset, and plot the learned weights distribution \wrt clean and noisy interactions as well as their loss values in Figure~\ref{fig:visualization}. We find that:
\begin{itemize}[leftmargin=*]
    \item  The left of Figure~\ref{fig:visualization} indicates that almost all large weights belong to clean interactions, and the weights of noisy interactions are much smaller than that of clean ones,
    meaning that the SGDL can differentiate clean implicit feedback from the noisy one. %and noisy implicit feedback.
    \item The learned weighting function of SGDL in the right of Figure~\ref{fig:visualization} demonstrates that when the loss has relatively small values, the weighting function inclines to increase the weight together with loss, indicating that it tends to emphasize more on clean data for learning user preference; while as the loss gradually becomes larger, the weighting function first remains unchanged and then begins to dramatically decrease its weight, implying that it tends to highlight the hard yet clean interactions with large weights first and then suppress noise interactions. Therefore, SGDL is able to automatically locate hard interactions for better learning user preferences.
\end{itemize}

\section{Related Work}
\label{sec:related_work}

Existing recommender systems are typically trained with implicit feedback. Recently, some studies~\cite{sigir20model,wsdm21denoise,mm21occf,sigir21boostrapping} have noticed that implicit feedback could be easily corrupted by different factors (\eg popularity bias~\cite{arxiv20debias-survey} and unawareness of users' behaviors~\cite{cidm08noise}), and the inevitable noise would dramatically degrade the recommendation performance~\cite{mm21occf,wsdm21denoise,arxiv22ensemble,arxiv21ot}. As a result, some efforts have been dedicated to solving the noisy implicit feedback problem, which can be categorized into sample selection methods~\cite{kdd12ranking,www19adversarial,tkde19sampling-bpr,sigir20sampler,mm21occf} and sample re-weighting methods~\cite{wsdm21denoise,cikm21sequential-denoise,arxiv22ensemble}.

\vspace{5pt}
\noindent
\textbf{Sample Selection.} A simple idea to denoise implicit feedback is to select clean and informative samples only, and train the recommendation model with them. For example, WBPR~\cite{kdd12ranking} considers that the missing interactions of popular items are highly likely to be real negative examples, and hence assigns higher sampling probabilities to them. 
%associates them with higher probability to be sampled.
IR~\cite{mm21occf} interactively generates pseudo-labels for user preferences based on the difference between labels and predictions, to discover the noisy-positive and noisy-negative examples. Nonetheless, their performance has high variance since they heavily depend on the sampling distribution~\cite{uai18fbgd}.

\vspace{5pt}
\noindent
\textbf{Sample Re-weighting.} On the other hand, loss-based methods focus on the learning process of models (\eg loss values and predictions) to distinguish noisy interactions from clean data. For instance, T-CE~\cite{wsdm21denoise} dynamically assigns lower weights to high-loss samples since it has been shown that noisy examples would have larger loss values. DeCA~\cite{arxiv22ensemble} develops an ensemble method to minimize the KG-divergence between the two models' predictions, under the assumption that different models make relatively similar predictions on clean examples. Although these methods achieve promising results without additional data, they heavily rely on the predefined loss function and hyperparameters, incurring poor generalization for different recommendation models.

\vspace{5pt}
\noindent
\textbf{Other Directions.} There are some recent studies that consider using additional information~\cite{recsys14click,ijcai20feedback,recsys21denoise} or designing model-specific structures~\cite{sigir21enhanced,sigir21mask,sigir21sgl} to improve the robustness of recommender systems. For instance, DFN~\cite{ijcai20feedback} proposes a feedback interaction component to extract clean and useful information from noisy feedback with additional explicit feedback (\eg like and dislike). SGCN~\cite{sigir21mask} treat user-item interactions as a bipartite graph, and devise a learnable regularization module to preserve the sparsity and low rank of graph structure. SGL~\cite{sigir21sgl} advances graph-based recommender systems with self-supervised learning by employing graph structure augmentations. However, these methods suffer from poor generalization, since they either need additional information (\eg explicit feedback) as guidance to denoise implicit feedback~\cite{ijcai20feedback,recsys21denoise}, or are only applicable to specific data structures (\eg user-item bipartite graph)~\cite{sigir21mask,sigir21enhanced,sigir21sgl}.

\vspace{5pt}
\noindent
\textbf{Difference from Existing Work.} Our work can be seen as a variant of self-training~\cite{nips20self-training}, which leverages self-labeled memorized data to enhance the robustness of training process. Compared with existing work, SGDL is in an orthogonal direction, opening up a new research line of denoising implicit feedback for recommendations. The most relevant work with ours is MORPH~\cite{kdd21self-trainsition}. However, it tackles the classification problem in computer vision, and only uses the memorized data for training, which is not feasible in ranking-based recommendations since the hard yet clean data may not be memorized without clear guidance. Compared with previous methods, SGDL is a truly model-agnostic framework, which can be easily applied to \textit{any} learning-based recommendation models with \textit{any} ranking loss functions, and does not need to define \textit{any} weighting functions.

% \vspace{-0.13in}
\section{Conclusion and Future Work}
\label{sec:conclusions}

In this paper, we present a new denoising paradigm, called SGDL, which leverages the memorization effect of recommendation models and designs two training phases to exploit the self-labeled memorization data as guidance for denoising learning. Extensive experiments conducted on three real-world datasets with four representative recommendation models demonstrate the superiority and universality of SGDL. In the future, we plan to jointly explore the noise and bias existing in implicit feedback to develop a universal denoising and debiasing solution. 

% \noindent
% \textbf{Acknowledgement.} This work was supported by the NSFC under Grants No. (62025206, 61972338, and 62102351).

%%
%% The acknowledgments section is defined using the "acks" environment
%% (and NOT an unnumbered section). This ensures the proper
%% identification of the section in the article metadata, and the
%% consistent spelling of the heading.

\begin{acks}
This work was supported by the NSFC under Grants No. (62025206, 61972338, and 62102351). 
Lu Chen is the corresponding author of the work.
\end{acks}

%%
%% The next two lines define the bibliography style to be used, and
%% the bibliography file.
\bibliographystyle{ACM-Reference-Format}
\balance

\bibliography{refer}
\balance

\end{document}